\documentstyle[aps,psfig,amssymb]{revtex}	

\begin{document}

\title{
\rightline{\small{\sl Twistor Newsletter\/} {\bf 43}, 
	      pp. 14-21 (1997)}
        Singularities of wavefronts and lightcones in the context 
	of GR via null foliations	
        }

\author{
	Simonetta Frittelli\thanks{e-mail: 
				simo@artemis.phyast.pitt.edu}$^{a,b}$
	\and
	Ezra T. Newman$^a$ 
	}
\address{
$^a$	 Department of Physics and Astronomy,
         University of Pittsburgh,
         Pittsburgh, PA 15260,
	 USA.							\\
$^b$	Department of Physics,
	Duquesne University,
	Pittsburgh, PA 15282
        }
\date{September 4, 1997}
\maketitle

\begin{abstract}

We describe an approach to the issue of the singularities of null
hypersurfaces, due to the focusing of null geodesics, in the context of
the recently introduced formulation of GR via null foliations.

\end{abstract}

\section*{}

The null-surface approach to general relativity essentially
reformulates general relativity in terms of two real functions on the
bundle of null directions over the spacetime manifold (locally
$M^4\times S^2$ with a metric $g_{ab}(x^a)$ on $M^4$).  These two
functions are $Z(x^a,\zeta,\bar\zeta)$, representing a sphere's worth
of null foliations of the spacetime (i.e., such that
$\tilde{g}^{ab}Z,_aZ,_b=0$ for all values of $\zeta$ and for members
$\tilde{g}_{ab}$ of the conformal class of $g_{ab}$), and
$\Omega(x^a,\zeta,\bar\zeta)$, representing a sphere's worth of
conformal factors with the role of picking the members of the the
conformal class that satisfy the Einstein equations.  More detail on
this formulation can be found in~\cite{fkn95b}.

Within the context of the null-surface approach to general relativity a
family of null coordinate systems $(\theta^i, i=0,1,+,-)$ is heavily
used to derive dynamical equations for $Z$ and $\Omega$.  The
coordinates $\theta^i$ are defined by derivation from $Z$:
\begin{equation}
	\theta^i = (Z, \eth\bar\eth Z, \eth Z, \bar\eth Z)
		\equiv (u,R,\omega,\bar\omega),
							\label{coord}
\end{equation}
which, for fixed $\zeta$, represents a transformation between an
arbitrary coordinate system $x^a$ and $\theta^i$.  The coordinates
$\theta^i$ are adapted to the null foliations, so that the leaves
$Z(x^a,\zeta,\bar\zeta)=const.$ of the foliation constitute surfaces of
constant coordinate $u$.

There is some gauge freedom in the theory, due to the fact that there
are different foliations which are all null with respect to the same
metric, such as, for Minkowski spacetime, a foliation based on outgoing 
lightcones off a world
line as opposed to a foliation by null planes. In the case of
asymptotically flat spacetimes, we customarily fix the gauge by
requiring that our null foliation consists of surfaces that
asymptotically become null planes.

For every fixed value of $\zeta$, our special coordinates have a well
defined interpretation.  The coordinate $u$ labels leaves of the null
foliation.  The coordinates $(\omega,\bar\omega)$ label null geodesics
on a fixed leaf.  The remaining coordinate $R$ acts as a parameter
along every null geodesic in the leaf.   Because of our gauge choice of
null surfaces becoming planes at null infinity, the null geodesics
labeled by $(\omega,\bar\omega)$ constitute bundles of asymptotically
parallel null geodesics.

It is natural to raise the objection that, generically, any vacuum
spacetime other than Minkowski focuses non-diverging bundles of null
geodesics~\cite{hawking}, and therefore our coordinate systems break
down at the point of focusing, by assigning different labels to the
same spacetime point.  In this respect, although coordinate
singularities are irrelevant to the physical content of the dynamical
null-surface equations,  we feel that the break down of these
particular coordinates has a certain appeal since it entails the
existence and location of caustics.  Caustics and their singularities,
as well as the singularities of wavefronts, have been classified by
Arnol'd within a sophisticated mathematical
context~\cite{arnold}. On the other hand, caustics are increasingly
being considered in the field of astrophysical 
observations~\cite{ellis,petters93,petters97,GL}.

The null-surface approach provides a dual interpretation to the
function $Z$.  The condition $Z(x^a,\zeta,\bar\zeta)=u$ for fixed $x^a$
picks up the points $(u,\zeta,\bar\zeta)$ at scri which are connected
to $x^a$ by null geodesics.  These points lie on a two-surface at scri,
referred to as the lightcone cut of the point $x^a$.  Generically, due
to focusing in the interior, the lightcone cuts have self-intersections
and typical wavefront singularities such as cusps and swallowtails.
This appears to pose a technical difficulty regarding the perturbative
approach to solving the null-surface equations, since the occurrence of
this type of singularity generically entails divergences in the
derivatives of the lightcone cuts.  

In the following, we examine the occurrence of singularities of
wavefronts and lightcones and its relevance to the null-surface
approach.

\subsection{Singularities of the null coordinates}

\begin{figure}
\centerline{
		\psfig{figure=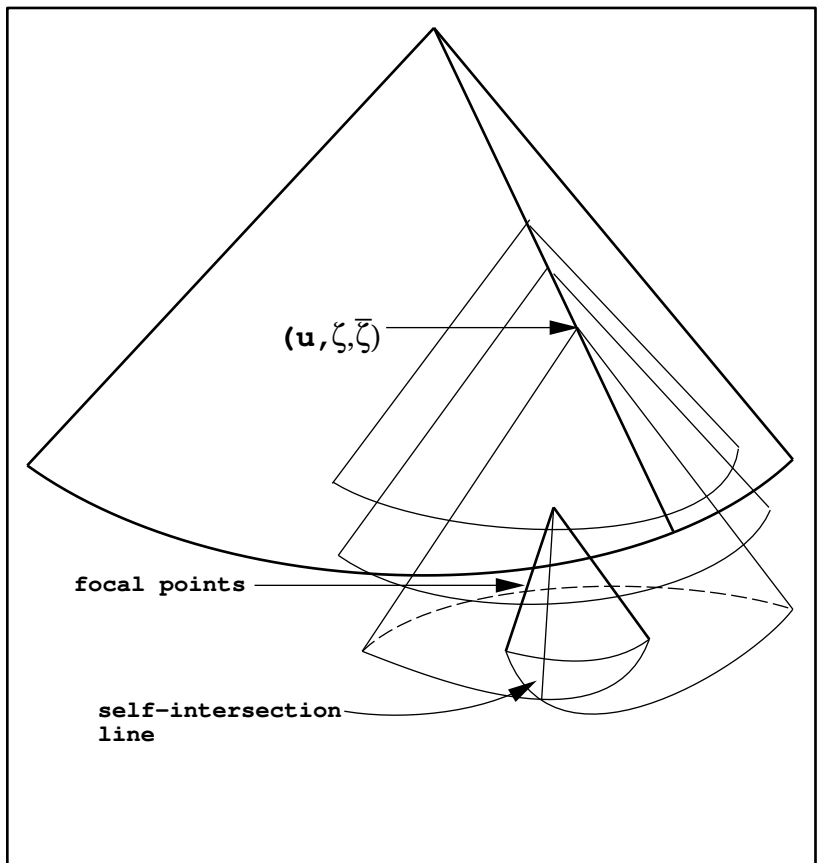,height=4.5in,angle=0}
}
		\caption{A null foliation by past lightcones of 
			 points at scri.}
		\label{parallel}
\end{figure}

Consider a foliation of spacetime by past null cones from points
$(u,\zeta,\bar\zeta)$ at scri along a fixed null generator $\zeta$.
(This is the compactified version of a foliation by null surfaces that
are asymptotically null planes).  Every past lightcone in this
foliation has singularities, in the sense that the lightcone ``folds''
and self-intersects, due to focusing in the interior spacetime, as
shown in Fig. \ref{parallel}.  The points where the past lightcone is
singular are points where neighboring null geodesics of the congruence
intersect, and are thus conjugate to the point at scri.  Translating
this into the physical non-compactified spacetime,
these points in the interior are ``focal points'', such that light rays
emitted from them are asymptotically parallel.  How can we locate these
``focal points'' in terms of null-surface variables?


A preliminary answer to this question can be approached quite directly
from a consideration of the geodesic deviation vector of the congruence
that becomes asymptotically parallel in a direction $(\zeta,\bar\zeta)$
at scri.  Every null geodesic in this congruence is characterized by
fixed values of $(u,\zeta,\bar\zeta,\omega,\bar\omega)$.  The geodesic
deviation vector has been derived earlier in~\cite{geodesicdev}.  Along
a fixed null geodesic, the cross sectional area of the congruence has
the expression:
\begin{equation}
	A_p(R; u,\zeta,\bar\zeta,\omega,\bar\omega) 
    = 
	\frac{1}{\Omega^2\sqrt{1-\Lambda,_1\bar\Lambda,_1}},    \label{2}
\end{equation}
where $,_1$ represents $\partial/\partial R$.
This is an expression of the area $A_p$ in terms of the two
null-surface variables
\begin{equation}
	\Omega = \Omega(x^a,\zeta,\bar\zeta)\hspace{2cm}
	\Lambda\equiv \eth^2 Z(x^a,\zeta,\bar\zeta).
\end{equation}
In (\ref{2}), the quantities $\Omega(x^a,\zeta,\bar\zeta)$ and
$\Lambda(x^a,\zeta,\bar\zeta)$ are evaluated at fixed
$(\zeta,\bar\zeta)$ and at values of $x^a$ along the null geodesic
given by $(u,\zeta,\bar\zeta,\omega,\bar\omega)$.

It is relevant to point out that the variable $\Omega$ actually is the
product of two factors, one of which is an arbitrary confromal factor
for the conformal class (and as such, it does not depend on $\zeta$),
whereas the other factor carries conformal information via its
$\zeta$-dependence. Therefore it is not surprising to find that it
plays a role in a completely conformally invariant matter such as the
determination of conjugate points of null geodesic congruences.

Focusing takes place at the value of $R$ such that $A_p=0$.  However,
the only way for the area to vanish is that the denominator become
infinite.  The square root in the denominator can not diverge before
becoming pure imaginary.  This leaves us with the previously
unsuspected result that along a fixed null geodesic in the past
lightcone from a point $(u,\zeta,\bar\zeta)$ at scri, $\Omega$ must
blow up for focusing to take place.

The vanishing of $A_p$ is also related to the vanishing of the
determinant of the metric, since in these coordinates we have
\begin{equation}
	g \equiv \mbox{det}(g_{ij}) 
	       = \big( \mbox{det}(g^{ij}))^{-1}
	       = \frac{1}
		      {\Omega^8 (1-\Lambda,_1\bar\Lambda,_1)}
	       = \Omega^{-4}A_p^2
\end{equation}
Thus both the vanishing of $A_p$ and the fact that $\Omega$ diverges
result in the vanishing of the 4-dimensional volume element.

This divergence of $\Omega$ has another significant consequence.  Along
the null geodesics labeled by $(\omega,\bar\omega)$ there is a choice
of an affine parameter $s$, which is related to $R$ via
  \begin{equation}
	\frac{ds}{dR} = \Omega^{-2}
	\hspace{1cm}
	\mbox{or alternatively}
	\hspace{1cm}
	\frac{dR}{ds} = \Omega^{2}.
  \end{equation}
Since the affine parameter is regular, it follows that $dR/ds$ blows up
as well at the point where $\Omega$ does.  Thus $R$ is a bad coordinate
(as we might have suspected) in the neighborhood of a focal point.
Both $R$ and $\Omega$ are, however, determined by the function $Z$
through the same second derivative (See~\cite{fkn95b} for details):
  \begin{equation}
	R\equiv\eth\bar\eth Z(x^a,\zeta,\bar\zeta),
  \hspace{2cm}
	\Omega^2 \equiv g^{ab}Z,_a\eth\bar\eth Z,_b,
  \end{equation}
thus it is consistent to attribute both complications to $\eth\bar\eth
Z(x^a,\zeta,\bar\zeta)$ becoming singular, since $g^{ab}$ is smooth in
a good choice of coordinates $x^a$.

\subsection{Singularities of the lightcone cuts}

In asymptotically flat spacetimes the function $Z$ can also be viewed
as describing the intersection of the lightcone of a point $x^a$ with
scri, via $u=Z(x^a,\zeta,\bar\zeta)$ where $(u,\zeta,\bar\zeta)$ are
Bondi coordinates on scri. This intersection is a two-surface in a
three-dimensional space and can be thought as one member of a
series of ``wavefronts'' obtained by slicing the lightcone of
a point $x^a$ with a one-parameter family of past lightcones, the last
one of them being scri itself.  In this respect, the two-surface at
scri given by $u=Z(x^a,\zeta,\bar\zeta)$ can be thought of as a
two-dimensional wavefront in a three-dimensional space, for which there
is a standard treatment in singularity theory.

Wavefronts are considered as projections of smooth two-dimensional
Legendrian manifolds in a five-dimensional space down to a
three-dimensional space.  The singularities of the wavefront are the
places where the projection is singular.  The Legendrian manifold
itself is obtained from a generating function.  More specifically,
consider the 1-jet bundle over the two dimensional configuration space
$(x^1,x^2)$, given in coordinates $(z,x^1,x^2, p_1,p_2)$, and a
function $S(p_1,x^2)$. A Legendrian manifold is a two dimensional
subspace of points $(z,x^1,x^2, p_1,p_2)$ that can be specified
by~\cite{arnold}
\begin{equation}
	x^1 = \frac{\partial S}
		   {\partial p_1},	\hspace{1cm}
	p_2 = -\frac{\partial S}
		    {\partial x^2},	\hspace{1cm}
	z   = S(p_1,x^2) - x^1p_1,	\hspace{1cm}
	\mbox{ parametrized by $x^2$ and $p_1$.}
\end{equation}
The projection of this Legendrian manifold down to $(z,x^1,x^2)$ is a
two-surface in three dimensions, representing a wavefront.  This
wavefront is singular where the projection breaks down, namely at
points such that the $3\times 2$ Jacobian matrix
\begin{equation}
	\left(
	\begin{array}{cc}
	\displaystyle{\frac{\partial x^1}{\partial x^2}}	&
	\displaystyle{\frac{\partial x^1}{\partial p_1}}	\\
								\\
	\displaystyle{\frac{\partial x^2}{\partial x^2}}	&
	\displaystyle{\frac{\partial x^2}{\partial p_1}}	\\
								\\
	\displaystyle{\frac{\partial z  }{\partial x^2}}	&
	\displaystyle{\frac{\partial z  }{\partial p_1}}	
	\end{array}
	\right)
\end{equation}
has rank less than 2.  

In our case, we take $(x^1,x^2)$ as coordinates on the sphere by
defining $\zeta = x^1+ix^2$ and interpret the projection of $z$ as the
lightcone cut function $Z(x^a,\zeta,\bar\zeta)$ (the dependence on the
spacetime points $x^a$ is considered parametric in the instance of
lightcone cuts, being regarded as fixed here).  The lightcone cut is
regular except at some values of $\zeta$ at which the projection breaks
down. Carrying through the calculation of the determinant of the
Jacobian matrix in these terms, we obtain the result that the
projection breaks down at points $\zeta$ such that
\begin{equation}
	\frac{1}{\eth^2 Z(x^a,\zeta,\bar\zeta)} = 0	\hspace{1cm}
	\mbox{and}					\hspace{1cm}
	\frac{1}{\eth\bar\eth Z(x^a,\zeta,\bar\zeta)} = 0.
\end{equation} 
This calculation is parametric in $x^a$, where $x^a$ represents the
apex of the lightcone intersecting scri at the lightcone cut surface.
This is equivalent to the statement that both $\Lambda$ and $R$ must
blow up at particular values of $\zeta$ for the lightcone cut of a
given spacetime point $x^a$ (See Fig.\ref{lightcone}). For every point
on the cut, the quantities $\Lambda$ and $R$ have finite values, except
at the singular points shown in Fig.\ref{lightcone}.  Given the
lightcone cut function $Z(x^a,\zeta,\bar\zeta)$ for fixed $x^a$, the
values of $\zeta$ for which $\Lambda$ and $R$ diverge determine the
null geodesics for which $x^a$ is a focal point. This is consistent
with the result found in the previous section.

\subsection{Singularities of the lightcones}

\begin{figure}
\centerline{
		\psfig{figure=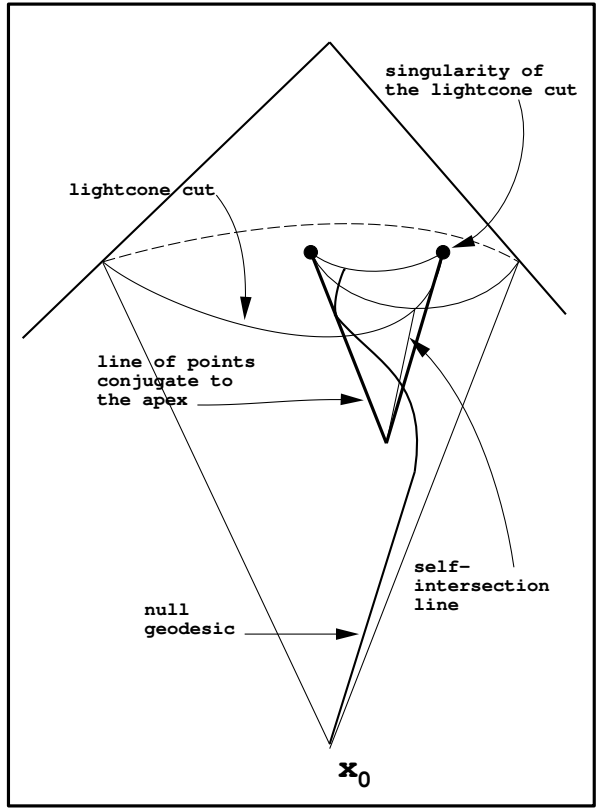,height=4.2in,angle=0}
}
		\caption{The lightcone and lightcone cut of a point $x^a_0$.
For the null geodesic shown, both $\Lambda(x^a_0,\zeta,\bar\zeta)$ and 
$R(x^a_0,\zeta,\bar\zeta)$ are finite. }
		\label{lightcone}

\end{figure}

A related issue which we are also concerned with is the location of the
singularities of the lightcone of a point in the interior spacetime.
Consider the lightcone of a point, namely, the congruence of all the
null geodesics through that point, and follow one null geodesic in the
congruence out to the future.  Generically, due to curvature, there is
at least one future point along this null geodesic where neighboring
geodesics intersect, referred to as a point conjugate to the apex.  The
cross sectional area of the congruence vanishes at points which are
conjugate to the apex along any null geodesic of the congruence, the
locust of all such points being sometimes referred to as the caustic
surface (although it would be more accurate to call it the singularity
of the lightcone).  The lightcone folds and self-intersects beyond the
occurrence of the earliest of the points conjugate to the apex, as shown
in Fig. \ref{lightcone}.  How do we formulate the condition for the
location of such singular points in terms of null-surface variables?

\begin{figure}
\centerline{
		\psfig{figure=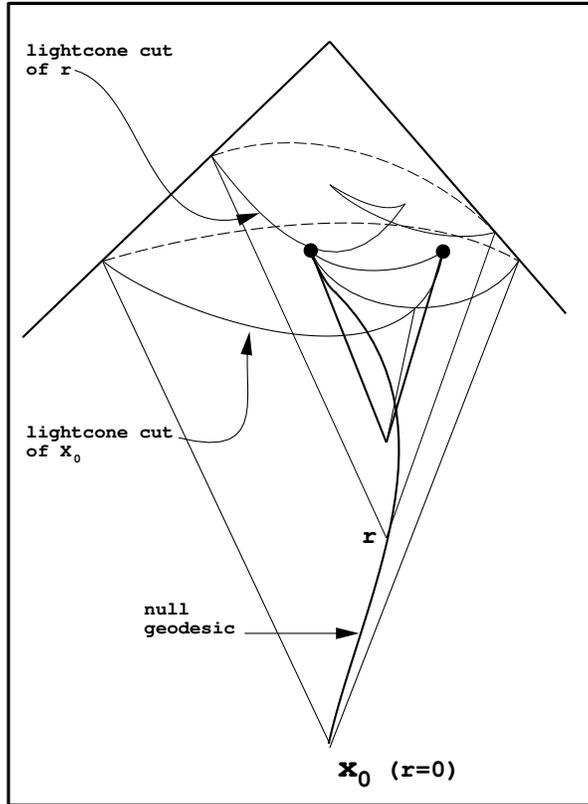,height=4.2in,angle=0}
}
		\caption{Evaluating $\Lambda$ and $R$ at points $s$ along 
the null geodesic. In this case, $\Lambda_0$ and $R_0$ blow up, where $\Lambda$
and $R$ are finite at the point $r$ further up, since the null geodesic 
does not hit a singularity of the lightcone cut of the point $r$. }
		\label{finitelambda}

\end{figure}
\begin{figure}
\centerline{
		\psfig{figure=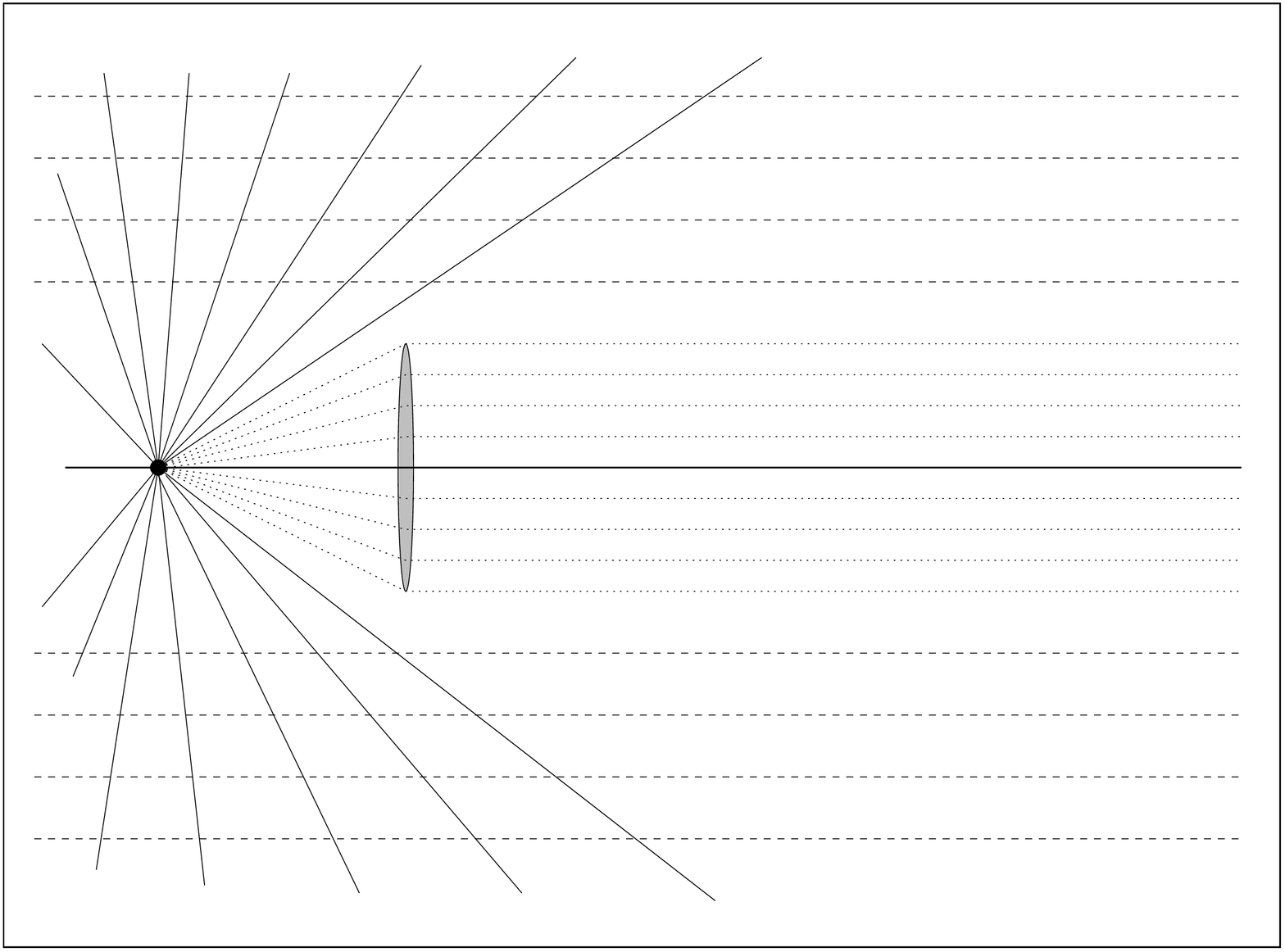,height=3in,angle=-90}
}
		\caption{Two null congruences containing the same null ray. 
The common null ray is shown in boldface.  The solid lines represent 
the lightcone congruence, whereas the dashed lines represent the congruence 
asymptotically parallel.  The dotted rays are also common to both 
congruences. In this case, the apex is a focal point. }
		\label{congruence3}

\end{figure}
\begin{figure}
\centerline{
		\psfig{figure=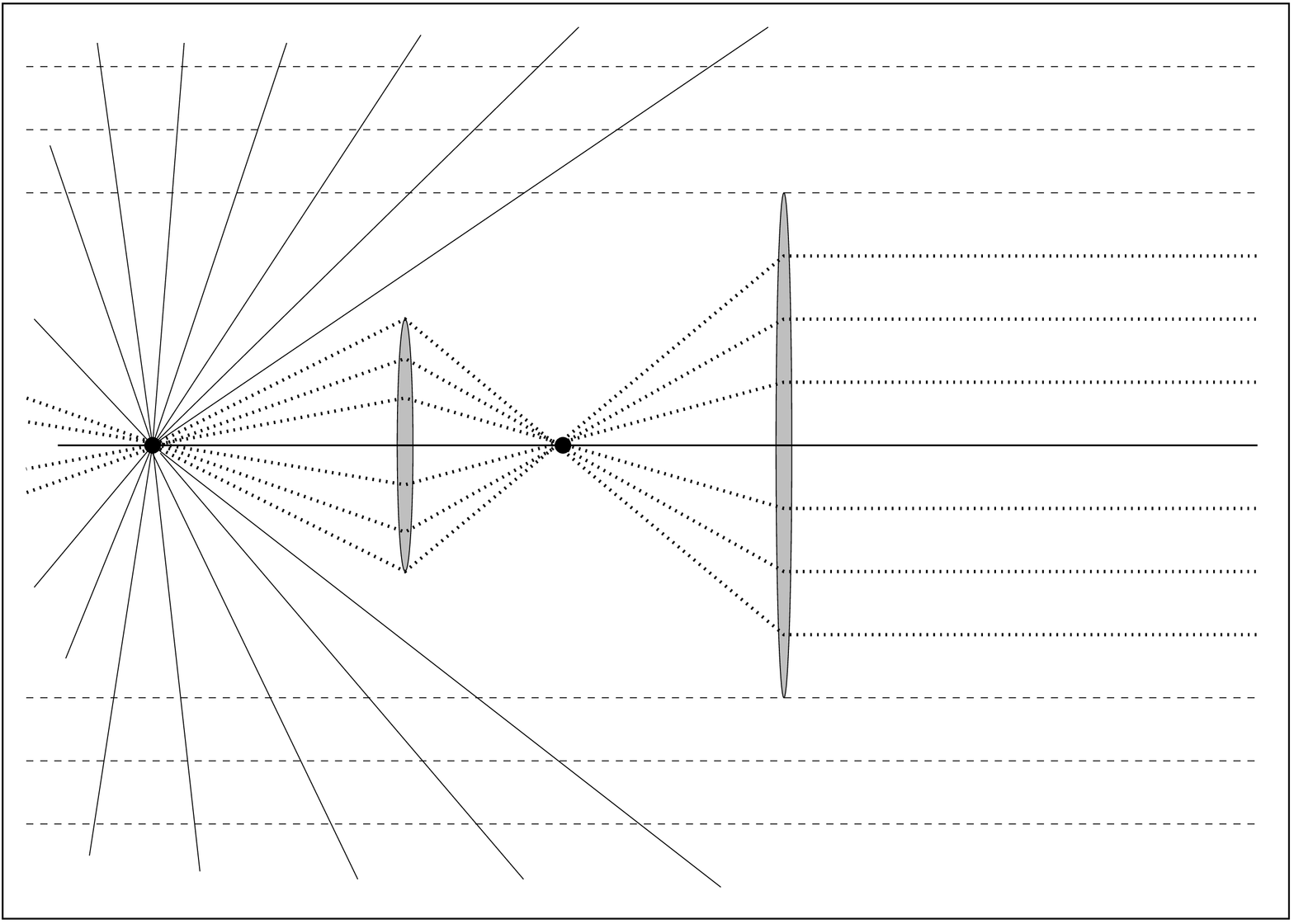,height=3in,angle=-90}
}
		\caption{Two null congruences containing the same null ray. 
The common null ray is shown in boldface.  The solid lines represent 
the lightcone congruence, whereas the dashed lines represent the congruence 
asymptotically parallel.  The dotted rays are also common to both 
congruences. In this case, the apex is a focal point and so is the 
focusing point between the two lenses. }
		\label{congruences}

\end{figure}

We can approach this issue from an analysis of the geodesic deviation
vector along a fixed null ray $\zeta$ of the lightcone of an interior
point $x_0^a$.  The geodesic deviation vector in this case can be
derived solely from knowledge of $Z(x^a,\zeta,\bar\zeta)$ through a
procedure that is standard to the null-surface approach.  The
coordinate transformation (\ref{coord}) can, in principle, be
piece-wise inverted, yielding, perhaps with different branches,
\begin{equation}
	x^a = f^a(u,\omega,\bar\omega,R, \zeta,\bar\zeta)	\label{cone}
\end{equation}
for fixed $\zeta$.  The lightcone of a point $x_0^a$ can be obtained
now by substituting
\begin{equation}
	     u =          Z(x_0^a,\zeta,\bar\zeta)	\hspace{1cm}
	\omega = \eth     Z(x_0^a,\zeta,\bar\zeta)	\hspace{1cm}
    \bar\omega = \bar\eth Z(x_0^a,\zeta,\bar\zeta)
\end{equation}
and 
\begin{equation}
	R = \eth\bar\eth Z(x_0^a,\zeta,\bar\zeta) + r
\end{equation}
into (\ref{cone}), in this manner obtaining, perhaps with several
branches,
\begin{equation}
	x^a = F^a(r; x_0^a,\zeta,\bar\zeta)
\end{equation}
as the lightcone of the point $x_0^a$ parametrized by the directions
$(\zeta,\bar\zeta)$ labeling the null geodesics, and the parameter $r$
along each null geodesic.  The geodesic deviation vector is
\begin{equation}
	M^a = \eth F^a = 
	\frac{\partial f^a}{\partial \theta^i} \eth\theta^i + \eth'f^a
\end{equation}
where $\eth'$ is taken keeping $\theta^i$ fixed.   This expression can
be worked out straightforwardly to yield $M^a$ in terms of $Z$ and its
derivatives, and the cross sectional area of the congruence along the
null ray $\zeta$ can subsequently be obtained:
\begin{equation}
	A_{lc}(r;x_0^a,\zeta,\bar\zeta)
   =
	 \frac{(\Lambda     -     \Lambda_0)
	       (\bar\Lambda - \bar\Lambda_0)
			    - (R - R_0)^2}
		{\Omega^2 \sqrt{1 - 
		\Lambda,_1
	    \bar\Lambda,_1}}				\label{alc}
\end{equation} 
The sublabel $0$ indicates evaluation at $r=0$ (the apex).  The
quantities $\Lambda_0$ and $R_0$ appearing in (\ref{alc}) can be
thought of as referring to the lightcone cut of the apex, whereas
$\Lambda$ and $R$ correspond to the lightcone cut of the point at $r$
along the null geodesic labeled by $\zeta$ (See Fig.
\ref{finitelambda}).  By comparing with (\ref{2}) we can see that
\begin{equation}
	A_{lc} = A_p \; H(r;x_0^a,\zeta,\bar\zeta)
\hspace{1cm}\mbox{with}\hspace{1cm}
	H(r;x_0^a,\zeta,\bar\zeta)
		\equiv (\Lambda - \Lambda_0)
			(\bar\Lambda - \bar\Lambda_0)
			- (R - R_0)^2.
\end{equation} 
This equation relates the cross sectional areas of two different
congruences containing the same null ray $\zeta$, namely the cone of
lightrays through $x^a_0$ and the congruence of asymptotically parallel
rays parallel to the null ray $\zeta$.  This implies a relationship
between the focusing of either congruence and the behavior of the
quantities $\Lambda$ and $R$.  We distinguish three alternatives.
  \begin{enumerate}
     \item $A_{lc}=0$ and $A_p=0$ at some value $r$. 
	Then $H$ must not diverge at a rate faster than $A_p^{-1}$ 
	at that point.
     \item $A_{lc}=0$ with $A_p \neq 0$ at some value $r$. 
	Then $H$ must vanish at that point.
     \item $A_p = 0$ with $A_{lc}\neq 0$ at some value $r$. 
	Then $H$ must blow up at a rate faster than  $A_p^{-1}$ 
	at that point.
\end{enumerate}
As an example of the first instance, if $A_p=0$ at the apex then the
apex is a focal point, because $H=0$, and is therefore finite, at $r=0$
(See Fig~\ref{congruence3}).  So should be any point conjugate to it
along the null ray $\zeta$; however, at such points $\Lambda_0, R_0,
\Lambda$ and $R$ all blow up, therefore $H$ becomes quite intractable
and its behavior has yet to be verified (See Fig. \ref{congruences}).

\begin{figure}
\centerline{
		\psfig{figure=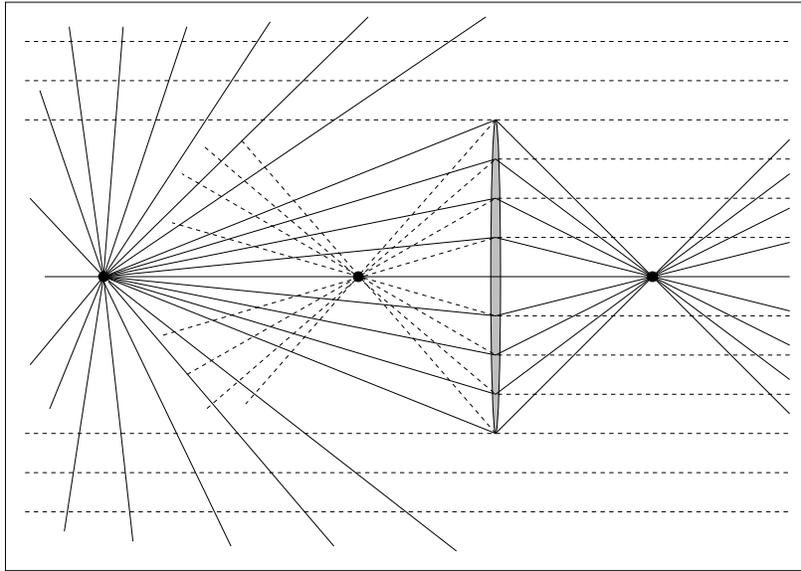,height=3in,angle=-90}
}
		\caption{Two null congruences containing the same null ray. 
The common null ray is shown in boldface.  The solid lines represent the 
lightcone congruence, whereas the dashed lines represent the congruence 
asymptotically parallel. In this case, the apex is not a focal point, 
nor is its conjugate point beyond the lens. The focal point is between the
lens and the apex. }
		\label{congruences2}

\end{figure}

In the second case, the apex is not a focal point, and its conjugate
point is located at $r$ such that 
\begin{equation}
	H(r;x_0^a,\zeta,\bar\zeta)=0.
\end{equation}
This is the generic situation illustrated in Fig.\ref{lightcone}. See also
Fig. \ref{congruences2}.

In the third case, the point $r$ at which $A_p$ vanishes is a focal
point along the null ray $\zeta$, and therefore $\Lambda$ and $R$ both
blow up, thus again $H$ becomes intractable but is not unlikely that it
blows up, since $\Lambda_0$ and $R_0$ are finite in this case. See 
Fig. \ref{congruences2}.

This analysis applies to every null ray $\zeta$ in the lightcone.  For
some null ray $\zeta_m$ the conjugate point occurs closest to the apex,
for some other direction $\zeta_f$ the apex has a conjugate point at
infinity, and finally there are values of $\zeta$ for which no focusing
takes place at any value of $r$, as can be seen from
Fig.\ref{lightcone}.

\subsection{Conclusion}

Although the work reported on here is very much in progress, we believe
we are finding significant clues as to what consequences the occurrence
of focusing of null geodesics has for the null-surface formulation of
general relativity.  Our ultimate goal is to find a way to integrate
the singularity issue with the null-surface dynamical equations, which
played no role in this discussion.

This work has been supported by the NSF under grant No. PHY 92-05109.



\end{document}